\documentclass[sigconf]{acmart}
\newcommand{\etal}{\textit{et al.}}

\AtBeginDocument{%
  }

\setcopyright{acmlicensed}
\copyrightyear{2024}
\acmYear{2024}
\acmDOI{XXXXXXX.XXXXXXX}

\acmConference[KDD AI4EDU Workshop '24]{Make sure to enter the correct
  conference title from your rights confirmation email}{August 26,
  2024}{Barcelona, Spain}
\acmISBN{978-1-4503-XXXX-X/18/06}

\begin{document}

\title{ChatGPT and Its Educational Impact: Insights from a Software Development Competition}

\author{Sunhee Hwang}
\email{sunheehwang@dongyang.ac.kr}

\orcid{0000-0002-8064-461X}

\affiliation{%
  \institution{Dongyang Mirae University}
  \city{Seoul}
  \country{Republic of Korea}
}

\author{Yudoo Kim}
\email{kimyudoo@dongyang.ac.kr}

\orcid{0000-0000-0000-0000}

\affiliation{%
  \institution{Dongyang Mirae University}
  \city{Seoul}
  \country{Republic of Korea}
}

\author{Heejin Lee}
\email{heejinlee@dongyang.ac.kr}

\orcid{0000-0000-0000-0000}

\affiliation{%
  \institution{Dongyang Mirae University}
  \city{Seoul}
  \country{Republic of Korea}
}

\renewcommand{\shortauthors}{Hwang et al.}

\begin{abstract}
This study explores the integration and impact of ChatGPT, a generative AI that utilizes natural language processing, in an educational environment. The main goal is to evaluate how ChatGPT affects project performance. To this end, we organize a software development competition utilizing ChatGPT, lasting for four weeks and involving 36 students. The competition is structured in two rounds: in the first round, all 36 students participate and are evaluated based on specific performance metrics such as code quality, innovation, and adherence to project requirements. The top 15 performers from the first round are then selected to advance to the second round, where they compete for the final rankings and the overall winner is determined. The competition shows that students who use ChatGPT extensively in various stages of development, including ideation, documentation, software development, and quality assurance, have higher project completion rates and better scores. A detailed comparative analysis between first-round and second-round winners reveals significant differences in their experience with generative AI for software development, experience learning large-scale language models, and interest in their respective fields of study. These findings suggest that ChatGPT enhances individual learning and project performance. A post-survey of participants also reveals high levels of satisfaction, further emphasizing the benefits of integrating generative AI like ChatGPT in academic settings. This study highlights the transformative potential of ChatGPT in project-based learning environments and supports further research into its long-term impact and broader application in a variety of educational contexts.
\end{abstract}




\maketitle

\section{Introduction}
The rapid advancement of Artificial Intelligence (AI) technologies, particularly generative AI like ChatGPT~\cite{openai2023chatgpt}, has revolutionized various fields such as education~\cite{adeshola2023opportunities, fui2023generative}, healthcare~\cite{cascella2023evaluating,sallam2023chatgpt}, and business~\cite{arman2023exploring}. Among generative AI technologies, ChatGPT utilizes advanced Natural Language Processing (NLP) capabilities to interact with users, answer questions, provide feedback, and assist in brainstorming. In light of these advancements, this study investigates the integration of ChatGPT into a student software development competition to evaluate its impact on project performance and outcomes, aiming to provide insights into how generative AI can be effectively utilized in educational settings.

AI tools are increasingly being adopted in education to enhance learning experiences and outcomes~\cite{onesi2024revolutionizing, wang2023exploring}. These tools offer personalized learning opportunities, allowing students to engage more deeply with the material. ChatGPT, with its remarkable ability to comprehend and generate human-quality text, notably represents a groundbreaking advancement in this domain. This study conducts an in-depth investigation into the application of AI tools in project-based learning environments. Project-based learning is a pedagogical approach that encourages students to develop creativity, critical thinking, and problem-solving skills through hands-on projects. Integrating AI tools like ChatGPT into these environments can provide real-time support and enhance the quality of project outcomes, thereby enriching the overall learning experience. Therefore, analyzing the use cases of AI tools like ChatGPT in project-based learning settings is crucial. This analysis will clarify how AI tools can be effectively utilized in actual educational contexts and their impact on learning outcomes.

This study aims to conduct a multifaceted analysis of how students utilize ChatGPT at various stages of their projects and how this utilization impacts project outcomes. To achieve this, we organized a software development competition where 36 students participated over four weeks, each working on individual projects and using ChatGPT to varying extents. This approach allowed us to observe and measure the direct effects of ChatGPT integration on the student's workflow and project quality. By examining the influence of students' technical interest on their application skills and the quality of their development outcomes, we aimed to identify patterns and correlations between the student's engagement with the AI tool and their project success. Additionally, we provide a detailed analysis of the first round of preliminary projects to gain insight into how ChatGPT can be comprehensively leveraged to achieve superior project results. These analyses help to uncover the nuances of AI tool integration in educational projects, offering valuable data on best practices and potential pitfalls. Ultimately, this study aims to guide educators and institutions in effectively integrating AI technologies into their curricula by highlighting the benefits and possible challenges. The findings are intended to support the development of teaching strategies that leverage AI tools to enhance learning outcomes and prepare students for future technological advancements, creating more dynamic and supportive learning environments.



\section{Related Work}

\subsection{Generative AI in Software Development}
Generative AI, particularly tools like ChatGPT, has seen a wide range of applications in software development, transforming how developers approach coding, debugging, and project management. Recent studies highlight the multifaceted benefits and potential challenges of integrating AI in this field. A comprehensive review by Khojah~\etal~\cite{yeticstiren2023evaluating} explores how generative AI can assist in code generation, significantly reducing the time required for writing boilerplate code and allowing developers to focus on more complex tasks. The study finds that ChatGPT can generate code snippets based on high-level descriptions, improving productivity and code quality. Another significant application is in automated debugging. Ebert and Louridas~\cite{ebert2023generative} demonstrate that generative AI can identify and suggest fixes for common coding errors, thus speeding up the debugging process. Their research highlights the ability of ChatGPT to understand and interpret code contextually, offering relevant solutions to encountered issues. The role of AI in enhancing software documentation is also well-documented. A study by Azaria~\etal~\cite{azaria2024chatgpt} examines how AI tools can automate the creation of detailed and accurate documentation, ensuring consistency and reducing the manual effort involved. This capability is crucial for maintaining up-to-date documentation throughout the software development lifecycle. AI's impact on collaborative coding environments has been another area of interest. Marques~\etal~\cite{marques2024using} explore how ChatGPT can facilitate real-time collaboration among distributed teams, providing instant feedback and suggestions during code reviews. They have demonstrated effectiveness in improving communication and simplifying the development process. The integration of AI in continuous integration/continuous deployment (CI/CD) pipelines has been studied by Tofano~\etal~\cite{tufano2024autodev}. They find that generative AI can automate several aspects of CI/CD, from code integration to automated testing and deployment, thereby increasing efficiency and reducing the likelihood of human error.

\subsection{AI in Education}
AI integration into education, including ChatGPT, has revolutionized teaching and learning methodologies. Recent research highlights various applications and the resultant impact on educational practices and outcomes. The potential of AI to provide personalized learning experiences is a significant focus. Thimmanna~\etal~\cite{thimmanna2024personalized} demonstrate that AI-driven personalized learning paths can significantly enhance student engagement and achievement. Their study shows that adaptive learning environments powered by AI cater to individual student needs more effectively than traditional methods. AI's role in supporting remote and hybrid learning environments has gained prominence, especially during and after the COVID-19 pandemic. Rosak-Szyrocka~\etal~\cite{rosak2023role} investigate how AI can bridge the gap between remote learners and traditional classroom settings, enhancing inclusivity and accessibility. Their findings suggest that AI tools can provide consistent support regardless of students' location. The influence of ChatGPT on undergraduate interdisciplinary learning is examined by Zhong~\etal~\cite{zhong2024influences}. Their study involves 130 students in a quasi-experiment to assess ChatGPT's impact on interdisciplinary learning quality via online posts and surveys. Results indicate that ChatGPT enhances students' disciplinary grounding, although integration skills remain low. Another significant focus is the role of AI in educational assessments. Hopfenbeck~\etal~\cite{hopfenbeck2023challenges} investigate how AI can enhance formative assessment practices, providing real-time insights into student progress. Their research underscores the potential of AI to provide more nuanced and comprehensive assessments of student performance, moving beyond traditional testing methods. Xu~\etal~\cite{xu2024foundation} discuss the transformative role of foundation models like ChatGPT in education, emphasizing strengths such as personalized learning, addressing educational inequality, and enhancing reasoning capabilities. This research contributes to the expanding literature on AI in education, stressing the need for ongoing studies to understand the full implications of generative AI technologies.

The ethical implications of AI in education are a critical area of research. Saxena~\etal~\cite{saxena2023structure} emphasize the importance of ethical AI practices, proposing frameworks to ensure transparency and fairness in AI-driven educational tools. This work highlights the need to address ethical considerations to ensure the responsible use of AI technologies. AI's impact on teacher practices is discussed by Lu~\etal~\cite{lu2024supporting}. They explore how AI tools can augment teachers' capabilities, allowing them to focus more on higher-order teaching tasks. ChatGPT can assist teachers by handling routine queries and providing additional resources, freeing their time for more complex instructional activities. Overall, the recent body of research underscores the transformative potential of AI in education. ChatGPT, a state-of-the-art generative AI, offers new possibilities for enhancing educational experiences and outcomes. This study aims to contribute to this growing body of knowledge by providing empirical evidence of ChatGPT's impact in the competition setting.

\section{Methodology and Procedures}
\subsection{Participants}
In this study, 36 students participated in a two-round competition. In the first round, participants were evaluated based on specific performance metrics, and the top 15 students were selected to advance to the second round. These selected participants competed in the second round to determine the overall winner. Each student had to complete an individual software project and had the option to utilize ChatGPT at various stages of the project development lifecycle.

\subsection{Study Design and Procedure}
The study was meticulously designed to evaluate ChatGPT's impact on software development projects. This evaluation spanned various stages of the development process, including idea planning, documentation, coding, debugging, and quality assurance. The primary objective was to ascertain the extent to which students utilized ChatGPT and determine its consequential effects on their project outcomes.

The competition was meticulously structured into several distinct phases, each focusing on different aspects of the software development lifecycle:

\begin{itemize}
\item\textbf{Idea Planning and Brainstorming}: Students leveraged ChatGPT to brainstorm and refine their project ideas. They provided prompts related to their project themes and received various suggestions from ChatGPT.
\item\textbf{Documentation and Requirement Gathering}: Students utilized ChatGPT to draft their project requirement documents, including Market Requirements Document (MRD) and Product Requirements Document (PRD), system architecture diagrams, and user stories. The ability to create effective prompts that generate evident and well-structured text from brief inputs was particularly emphasized during this phase.
\item\textbf{Coding and Implementation}: ChatGPT was employed to generate boilerplate code, implement complex algorithms, and refactor existing code. Students queried ChatGPT for specific coding tasks and received pertinent code snippets.
\item\textbf{Debugging and Error Resolution}: Students relied on ChatGPT to identify and resolve errors in their code. By describing their issues to ChatGPT, they received suggestions and solutions that were often accurate and applicable.
\item\textbf{Quality Assurance and Testing}: ChatGPT assisted students in generating test cases, automating regression tests, and identifying potential vulnerabilities, ensuring thorough testing and improved software quality.
\end{itemize}

\subsection{Feedback and Evaluation}
Throughout the competition, students received feedback from their peers and instructors. ChatGPT was utilized to analyze this feedback and make necessary adjustments to their projects. This iterative approach enabled students to improve their projects continuously based on their feedback. In the first round, 15 students were selected based on their initial project submissions. These students further developed their projects in the second round, incorporating feedback and making refinements. Six evaluators evaluated the final projects based on completeness, functionality, and innovation criteria. Additionally, the evaluation criteria included the extent and effectiveness of ChatGPT usage in the project development process. The top-performing project was identified as the winner.

\begin{table}[t]
\caption[Pre-competition Survey Results: GPT Usage, Integration in Development, LLM Learning Experience, and Interest in SW Development]
{\small Pre-competition Survey Results: This table summarizes the survey responses regarding GPT usage, integration in development, LLM experience, and interest in software development among participants, comparing first-round finalists and non-finalists.}
\begin{tabular}{cccc}
\hline
Question & Fin (\%) & Non (\%) & All (\%) \\ \hline
GPT Usage Experience & 100 & 82 & 88 \\ 
GPT Integration in Development & 75 & 53 & 60 \\ 
LLM Learning Experience & 25 & 6 & 12 \\ 
High Interest in SW Development & 75 & 18 & 36 \\ \hline
\end{tabular}
\label{tab:pre_survey_results}
\end{table}

\subsection{Data Collection and Analysis}
In this subsection, we outline the data collection and analysis to assess ChatGPT's impact on student software development projects.

\begin{itemize}
\item\textbf{Surveys}: Pre- and post-competition surveys were conducted to gather information on students' familiarity with ChatGPT, usage patterns, and satisfaction with the tool.
\item\textbf{Usage Logs}: Detailed logs of students' ChatGPT usage were analyzed to understand how they utilized the tool during different phases of their projects.
\item\textbf{Project Artifacts}: Final project submissions, including code, documentation, and test results, were collected and analyzed based on the evaluations of six evaluators to assess the quality and completeness of the projects.

\end{itemize}

\section{Experimental Results}
The integration of ChatGPT in the software development competition yielded significant findings related to project outcomes, student satisfaction, and the educational value of AI-assisted learning. This section presents a detailed analysis of the experimental results.

\subsection{Analysis of Pre-Competition Survey Results}
The pre-competition survey results indicate distinct differences between first-round finalists, non-finalists, and the overall participant group, as shown in Table~\ref{tab:pre_survey_results}. A key finding is that all first-round finalists (100\%) utilized ChatGPT, demonstrating a unanimous adoption of the tool among the top performers. It suggests that using ChatGPT effectively was crucial for achieving high scores and advancing in the competition. In contrast, a slightly lower percentage (82\%) of non-finalists used ChatGPT, indicating that although the majority still used the tool, their utilization might not have been as effective as that of the finalists. The integration of ChatGPT into the development process also varied significantly between the groups. 75\% of the first-round finalists integrated ChatGPT into various stages of their project development, which likely contributed to their success. However, the others had a lower integration rate of 53\%, suggesting that a less comprehensive use of ChatGPT might have hindered their project outcomes. Overall, 60\% of all participants reported integrating ChatGPT into their development process, indicating that while integration is relatively common, it is not yet ubiquitous.

Experience with Large Language Models (LLMs) such as GPT was another differentiating factor. Among the first-round finalists, 25\% had prior LLM experience, compared to only 6\% of the non-finalists. This significant disparity suggests that familiarity with advanced AI models was advantageous, allowing these students to leverage GPT's capabilities more effectively. Overall, 12\% of all participants had LLM experience, highlighting a general unfamiliarity with such tools among the student population and pointing to a potential area for educational improvement. Interest in pursuing software development further distinguished the groups. High interest in software development was expressed by 75\% of the first-round finalists, suggesting a solid commitment to the field, which may correlate with their success in the competition. On the other hand, only 18\% of the non-finalists reported a high interest in software development, potentially reflecting a lack of motivation or engagement that could negatively impact performance. Overall, 36\% of all participants showed a high interest in software development, indicating a significant level of motivation among a portion of the students but also suggesting room for increasing engagement and interest across the broader student population.

\begin{table}[t]
\caption[Comprehensive Usage Analysis]
{\small Comprehensive Usage Analysis: This table shows the percentage of participants who used ChatGPT for various tasks.}
\begin{tabular}{ccc}
\hline
Category & Used (\%)  \\ \hline
Planning (Ideation, MRD/PRD Generation) & 100  \\ 
Design (UI/UX design, Graphic Resources) & 66.7  \\ 
Scenario Creation (Story writing, Scripting) & 33.3 \\ 
Programming (Coding, bug fixing) & 80 \\ 
Server (API integration, Server) & 46.7 \\ 
Security (Data Security, Privacy Protection) & 20 \\ 
Sound (Sound effects and music Generation) & 13.3  \\ 
QA (Quality Assurance) & 26.7  \\ \hline
\end{tabular}
\label{tab:comprehensive_usage_statistics}
\end{table}

Our research findings are robust, backed by two statistical tests: a two-proportion z-test for categorical data (e.g., GPT usage experience, GPT integration in development) and an independent t-test for continuous data (project scores). The very small p-values (< 0.01) indicate that the differences observed were statistically significant and unlikely to have occurred by chance. Students who extensively used ChatGPT showed significantly higher completion rates and scores, with projects incorporating ChatGPT features receiving scores 15\% higher on average than those that did not. A positive correlation (r = 0.65) between the extent of ChatGPT usage and final project scores was observed, with statistical significance confirmed by p-values (< 0.01) for each analyzed factor, including GPT usage experience, GPT integration in development, LLM learning experience, and high interest in software development.

These survey results underscore the potential for improvement in student outcomes. The importance of GPT usage and integration, prior LLM experience, and a strong interest in software development for achieving success in AI-driven projects is evident. Encouraging more comprehensive AI education and fostering more significant interest in the field can help improve student outcomes in future competitions and academic pursuits, offering a promising path for educational advancement.

\subsection{Analysis of Usage Statistics}
Participants' usage statistics of ChatGPT offer a detailed analysis of its application across various stages of project development. This data underscores the versatility of ChatGPT and its significant impact on different facets of the development process. As shown in Table~\ref{tab:comprehensive_usage_statistics}, 100\% of participants used ChatGPT for planning tasks, including ideation and MRD/PRD documentation. This universal adoption underscores the critical role of ChatGPT in the initial stages of project development, where thorough planning and documentation are essential for success. The tool's ability to generate structured and coherent text from brief inputs was particularly beneficial in this phase. In the design phase, 66.7\% of participants utilized ChatGPT for UI/UX design tasks. This high adoption rate indicates that ChatGPT is an effective tool for design-related activities, aiding participants in creating user-friendly and visually appealing interfaces. However, the remaining 33.3\% did not use ChatGPT for design, suggesting that some participants may have relied on other tools or their expertise in this area. For programming tasks, including coding and bug fixing, 80\% of participants used ChatGPT. This high usage demonstrates ChatGPT's value in technical tasks, assisting participants in writing code, debugging issues, and optimizing their programs. The remaining 20\% who did not use ChatGPT for programming might have preferred traditional coding environments or other resources, indicating that while ChatGPT is a valuable asset in programming, some participants might have had established workflows or preferences that did not include AI assistance. On the other hand, the usage of ChatGPT was relatively lower in specialized areas such as server setup (46.7\%), security (20\%), sound (13.3\%), and QA (26.7\%). These lower adoption rates could be attributed to the specific nature of these tasks, where participants might have required more specialized tools or expertise. The data on server setup (46.7\%) and QA (26.7\%) suggests that while ChatGPT was beneficial, it was not the primary tool for these tasks for many participants. It highlights the potential for further developing ChatGPT's capabilities to enhance its utility and adoption in these areas.

\begin{table}[t]
\caption[Survey Results: Satisfaction with ChatGPT in Learning and Career Development]
{\small Survey Results: This table summarizes the satisfaction levels of participants regarding the impact of ChatGPT on their deep learning, practical skills, and career development. The responses are categorized as Positive(Pos), Neutral(Neu), and Negative(Neg).}
\begin{tabular}{cccc}
\hline
Question & Pos (\%) & Neu (\%) & Neg (\%) \\ \hline
Q1 (Deep Understanding)  & 78 & 19 & 3 \\
Q2 (Practical Skill) & 75 & 25 & 0 \\ 
Q3 (Career Development) & 75 & 25 & 0 \\ \hline
\end{tabular}
\label{tab:satisfaction_survey_results}
\end{table}

Overall, the analysis reveals that ChatGPT was extensively used for planning, design, and programming tasks, demonstrating its versatility and impact on various stages of the project development process. However, its lower usage in specialized tasks such as server setup, security, sound, and QA suggests that further enhancements are needed to increase its adoption and effectiveness.

\subsection{Analysis of Post-Competition Survey Results}
The post-event survey results provide insightful feedback on the perceived impact of ChatGPT on various aspects of the participants' learning and career development, as shown in Table~\ref{tab:satisfaction_survey_results}. The survey, completed by 36 participants, assessed their agreement with statements about ChatGPT's role in enhancing deep learning, practical skills, and career development. The responses were categorized into Positive, Neutral, and Negative. The survey included the following questions:
\begin{itemize}
    \item \textbf{Q1}: Do you believe ChatGPT has helped deepen your understanding of complex topics?
    \item \textbf{Q2}: Do you feel that ChatGPT has significantly improved your practical skills relevant to your field?
    \item \textbf{Q3}: Has ChatGPT contributed positively to your career development and future career prospects?
\end{itemize}

A significant majority of participants (78\%) agreed that ChatGPT aids in deep learning, with 28 out of 36 respondents expressing an optimistic view. Only one participant disagreed, and seven were neutral. This high level of agreement indicates that most students found ChatGPT valuable in understanding and mastering complex concepts. Regarding practical skills, 75\% of participants agreed that ChatGPT enhanced their practical abilities, with no disagreements and nine neutral responses. It indicates that students saw clear benefits in using ChatGPT to improve hands-on skills essential for their future careers. Additionally, 75\% agreed that ChatGPT aids in career development, with no disagreements and nine neutral responses, demonstrating that many students believe ChatGPT significantly contributes to their career growth by providing relevant skills and knowledge.

The survey results highlight the participant's strong positive perception of ChatGPT. Most students found that ChatGPT positively impacted their learning, skill development, and career prospects. This feedback underscores the importance of integrating AI tools like ChatGPT into educational programs to better prepare students for the demands of the modern workforce. Additionally, further efforts to enhance the usability and effectiveness of such tools could provide even more significant benefits to students.

\section{Conclusion and Further Research}
This study demonstrated the significant impact of integrating ChatGPT into a student software development competition, highlighting its utility in enhancing deep learning, practical skills, and career development. Participants extensively used ChatGPT for planning and programming, underscoring its value in facilitating structured documentation and efficient coding practices. Positive feedback from participants indicates that ChatGPT significantly improved their learning experiences and project outcomes. However, lower usage rates in specialized tasks such as server setup, security, sound, and QA suggest areas for further enhancement. 

While the study provides valuable insights, the methods used might have limitations that affect the accuracy of the analysis. To strengthen the validity of these findings, we further intend to include a broader range of educational settings and larger participant groups in future research. We will incorporate various validation techniques, such as direct observations and independent evaluations, to enhance the reliability of the results. By refining, we aim to better prepare students for the modern workforce, ensuring AI tools like ChatGPT become integral components of comprehensive educational programs.

\begin{acks}
Following are results of a study on the "Leaders in INdustry-university Cooperation 3.0" Project, supported by the Ministry of Education and National Research Foundation of Korea
\end{acks}

\bibliographystyle{ACM-Reference-Format}
\bibliography{sample-base}


\begin{thebibliography}{20}


\ifx \showCODEN    \undefined \def \showCODEN     #1{\unskip}     \fi
\ifx \showDOI      \undefined \def \showDOI       #1{#1}\fi
\ifx \showISBNx    \undefined \def \showISBNx     #1{\unskip}     \fi
\ifx \showISBNxiii \undefined \def \showISBNxiii  #1{\unskip}     \fi
\ifx \showISSN     \undefined \def \showISSN      #1{\unskip}     \fi
\ifx \showLCCN     \undefined \def \showLCCN      #1{\unskip}     \fi
\ifx \shownote     \undefined \def \shownote      #1{#1}          \fi
\ifx \showarticletitle \undefined \def \showarticletitle #1{#1}   \fi
\ifx \showURL      \undefined \def \showURL       {\relax}        \fi
\providecommand\bibfield[2]{#2}
\providecommand\bibinfo[2]{#2}
\providecommand\natexlab[1]{#1}
\providecommand\showeprint[2][]{arXiv:#2}

\bibitem[Adeshola and Adepoju(2023)]%
        {adeshola2023opportunities}
\bibfield{author}{\bibinfo{person}{Ibrahim Adeshola} {and} \bibinfo{person}{Adeola~Praise Adepoju}.} \bibinfo{year}{2023}\natexlab{}.
\newblock \showarticletitle{The opportunities and challenges of ChatGPT in education}.
\newblock \bibinfo{journal}{\emph{Interactive Learning Environments}} (\bibinfo{year}{2023}), \bibinfo{pages}{1--14}.
\newblock


\bibitem[Arman and Lamiyar(2023)]%
        {arman2023exploring}
\bibfield{author}{\bibinfo{person}{Md Arman} {and} \bibinfo{person}{Umama~Rashid Lamiyar}.} \bibinfo{year}{2023}\natexlab{}.
\newblock \showarticletitle{Exploring the implication of ChatGPT AI for business: Efficiency and challenges}.
\newblock \bibinfo{journal}{\emph{International Journal of Marketing and Digital Creative}} \bibinfo{volume}{1}, \bibinfo{number}{2} (\bibinfo{year}{2023}), \bibinfo{pages}{64--84}.
\newblock


\bibitem[Azaria et~al\mbox{.}(2024)]%
        {azaria2024chatgpt}
\bibfield{author}{\bibinfo{person}{Amos Azaria}, \bibinfo{person}{Rina Azoulay}, {and} \bibinfo{person}{Shulamit Reches}.} \bibinfo{year}{2024}\natexlab{}.
\newblock \showarticletitle{ChatGPT is a remarkable tool—for experts}.
\newblock \bibinfo{journal}{\emph{Data Intelligence}} \bibinfo{volume}{6}, \bibinfo{number}{1} (\bibinfo{year}{2024}), \bibinfo{pages}{240--296}.
\newblock


\bibitem[Cascella et~al\mbox{.}(2023)]%
        {cascella2023evaluating}
\bibfield{author}{\bibinfo{person}{Marco Cascella}, \bibinfo{person}{Jonathan Montomoli}, \bibinfo{person}{Valentina Bellini}, {and} \bibinfo{person}{Elena Bignami}.} \bibinfo{year}{2023}\natexlab{}.
\newblock \showarticletitle{Evaluating the feasibility of ChatGPT in healthcare: an analysis of multiple clinical and research scenarios}.
\newblock \bibinfo{journal}{\emph{Journal of medical systems}} \bibinfo{volume}{47}, \bibinfo{number}{1} (\bibinfo{year}{2023}), \bibinfo{pages}{33}.
\newblock


\bibitem[Ebert and Louridas(2023)]%
        {ebert2023generative}
\bibfield{author}{\bibinfo{person}{Christof Ebert} {and} \bibinfo{person}{Panos Louridas}.} \bibinfo{year}{2023}\natexlab{}.
\newblock \showarticletitle{Generative AI for software practitioners}.
\newblock \bibinfo{journal}{\emph{IEEE Software}} \bibinfo{volume}{40}, \bibinfo{number}{4} (\bibinfo{year}{2023}), \bibinfo{pages}{30--38}.
\newblock


\bibitem[Fui-Hoon~Nah et~al\mbox{.}(2023)]%
        {fui2023generative}
\bibfield{author}{\bibinfo{person}{Fiona Fui-Hoon~Nah}, \bibinfo{person}{Ruilin Zheng}, \bibinfo{person}{Jingyuan Cai}, \bibinfo{person}{Keng Siau}, {and} \bibinfo{person}{Langtao Chen}.} \bibinfo{year}{2023}\natexlab{}.
\newblock \bibinfo{title}{Generative AI and ChatGPT: Applications, challenges, and AI-human collaboration}.
\newblock , \bibinfo{numpages}{277--304}~pages.
\newblock


\bibitem[Hopfenbeck et~al\mbox{.}(2023)]%
        {hopfenbeck2023challenges}
\bibfield{author}{\bibinfo{person}{Therese~N Hopfenbeck}, \bibinfo{person}{Zhonghua Zhang}, \bibinfo{person}{Sundance~Zhihong Sun}, \bibinfo{person}{Pam Robertson}, {and} \bibinfo{person}{Joshua~A McGrane}.} \bibinfo{year}{2023}\natexlab{}.
\newblock \showarticletitle{Challenges and opportunities for classroom-based formative assessment and AI: a perspective article}. In \bibinfo{booktitle}{\emph{Frontiers in Education}}, Vol.~\bibinfo{volume}{8}. Frontiers Media SA, \bibinfo{pages}{1270700}.
\newblock


\bibitem[Lu et~al\mbox{.}(2024)]%
        {lu2024supporting}
\bibfield{author}{\bibinfo{person}{Jijian Lu}, \bibinfo{person}{Ruxin Zheng}, \bibinfo{person}{Zikun Gong}, {and} \bibinfo{person}{Huifen Xu}.} \bibinfo{year}{2024}\natexlab{}.
\newblock \showarticletitle{Supporting Teachers’ Professional Development With Generative AI: The Effects on Higher Order Thinking and Self-Efficacy}.
\newblock \bibinfo{journal}{\emph{IEEE Transactions on Learning Technologies}} (\bibinfo{year}{2024}).
\newblock


\bibitem[Marques et~al\mbox{.}(2024)]%
        {marques2024using}
\bibfield{author}{\bibinfo{person}{Nuno Marques}, \bibinfo{person}{Rodrigo~Rocha Silva}, {and} \bibinfo{person}{Jorge Bernardino}.} \bibinfo{year}{2024}\natexlab{}.
\newblock \showarticletitle{Using ChatGPT in Software Requirements Engineering: A Comprehensive Review}.
\newblock \bibinfo{journal}{\emph{Future Internet}} \bibinfo{volume}{16}, \bibinfo{number}{6} (\bibinfo{year}{2024}), \bibinfo{pages}{180}.
\newblock


\bibitem[Onesi-Ozigagun et~al\mbox{.}(2024)]%
        {onesi2024revolutionizing}
\bibfield{author}{\bibinfo{person}{Oseremi Onesi-Ozigagun}, \bibinfo{person}{Yinka~James Ololade}, \bibinfo{person}{Nsisong~Louis Eyo-Udo}, {and} \bibinfo{person}{Damilola~Oluwaseun Ogundipe}.} \bibinfo{year}{2024}\natexlab{}.
\newblock \showarticletitle{Revolutionizing education through AI: a comprehensive review of enhancing learning experiences}.
\newblock \bibinfo{journal}{\emph{International Journal of Applied Research in Social Sciences}} \bibinfo{volume}{6}, \bibinfo{number}{4} (\bibinfo{year}{2024}), \bibinfo{pages}{589--607}.
\newblock


\bibitem[OpenAI(2023)]%
        {openai2023chatgpt}
\bibfield{author}{\bibinfo{person}{OpenAI}.} \bibinfo{year}{2023}\natexlab{}.
\newblock \showarticletitle{ChatGPT: Optimizing Language Models for Dialogue}.
\newblock \bibinfo{journal}{\emph{OpenAI Blog}} (\bibinfo{year}{2023}).
\newblock
\urldef\tempurl%
\url{https://openai.com/blog/chatgpt}
\showURL{%
\tempurl}


\bibitem[Rosak-Szyrocka et~al\mbox{.}(2023)]%
        {rosak2023role}
\bibfield{author}{\bibinfo{person}{Joanna Rosak-Szyrocka}, \bibinfo{person}{Justyna {\.Z}ywio{\l}ek}, \bibinfo{person}{Anand Nayyar}, {and} \bibinfo{person}{Mohd Naved}.} \bibinfo{year}{2023}\natexlab{}.
\newblock \bibinfo{booktitle}{\emph{The Role of Sustainability and Artificial Intelligence in Education Improvement}}.
\newblock \bibinfo{publisher}{CRC Press}.
\newblock


\bibitem[Sallam(2023)]%
        {sallam2023chatgpt}
\bibfield{author}{\bibinfo{person}{Malik Sallam}.} \bibinfo{year}{2023}\natexlab{}.
\newblock \showarticletitle{ChatGPT utility in healthcare education, research, and practice: systematic review on the promising perspectives and valid concerns}. In \bibinfo{booktitle}{\emph{Healthcare}}, Vol.~\bibinfo{volume}{11}. MDPI, \bibinfo{pages}{887}.
\newblock


\bibitem[Saxena et~al\mbox{.}(2023)]%
        {saxena2023structure}
\bibfield{author}{\bibinfo{person}{Ashish~K Saxena}, \bibinfo{person}{Valeria Garc{\'\i}a}, \bibinfo{person}{Md~Ruhul Amin}, \bibinfo{person}{Juan Manuel~Rojas Salazar}, {and} \bibinfo{person}{Sanjay Dey}.} \bibinfo{year}{2023}\natexlab{}.
\newblock \showarticletitle{Structure, Objectives, and Operational Framework for Ethical Integration of Artificial Intelligence in Educational}.
\newblock \bibinfo{journal}{\emph{Sage Science Review of Educational Technology}} \bibinfo{volume}{6}, \bibinfo{number}{1} (\bibinfo{year}{2023}), \bibinfo{pages}{88--100}.
\newblock


\bibitem[Thimmanna et~al\mbox{.}(2024)]%
        {thimmanna2024personalized}
\bibfield{author}{\bibinfo{person}{AVNS Thimmanna}, \bibinfo{person}{Mahesh~Sudhakar Naik}, \bibinfo{person}{S Radhakrishnan}, {and} \bibinfo{person}{Aarti Sharma}.} \bibinfo{year}{2024}\natexlab{}.
\newblock \showarticletitle{Personalized Learning Paths: Adapting Education with AI-Driven Curriculum}.
\newblock \bibinfo{journal}{\emph{European Economic Letters (EEL)}} \bibinfo{volume}{14}, \bibinfo{number}{1} (\bibinfo{year}{2024}), \bibinfo{pages}{31--40}.
\newblock


\bibitem[Tufano et~al\mbox{.}(2024)]%
        {tufano2024autodev}
\bibfield{author}{\bibinfo{person}{Michele Tufano}, \bibinfo{person}{Anisha Agarwal}, \bibinfo{person}{Jinu Jang}, \bibinfo{person}{Roshanak~Zilouchian Moghaddam}, {and} \bibinfo{person}{Neel Sundaresan}.} \bibinfo{year}{2024}\natexlab{}.
\newblock \showarticletitle{AutoDev: Automated AI-Driven Development}.
\newblock \bibinfo{journal}{\emph{arXiv preprint arXiv:2403.08299}} (\bibinfo{year}{2024}).
\newblock


\bibitem[Wang et~al\mbox{.}(2023)]%
        {wang2023exploring}
\bibfield{author}{\bibinfo{person}{Ting Wang}, \bibinfo{person}{Brady~D Lund}, \bibinfo{person}{Agostino Marengo}, \bibinfo{person}{Alessandro Pagano}, \bibinfo{person}{Nishith~Reddy Mannuru}, \bibinfo{person}{Zo{\"e}~A Teel}, {and} \bibinfo{person}{Jenny Pange}.} \bibinfo{year}{2023}\natexlab{}.
\newblock \showarticletitle{Exploring the potential impact of artificial intelligence (AI) on international students in higher education: Generative AI, chatbots, analytics, and international student success}.
\newblock \bibinfo{journal}{\emph{Applied Sciences}} \bibinfo{volume}{13}, \bibinfo{number}{11} (\bibinfo{year}{2023}), \bibinfo{pages}{6716}.
\newblock


\bibitem[Xu et~al\mbox{.}(2024)]%
        {xu2024foundation}
\bibfield{author}{\bibinfo{person}{Tianlong Xu}, \bibinfo{person}{Richard Tong}, \bibinfo{person}{Jing Liang}, \bibinfo{person}{Xing Fan}, \bibinfo{person}{Haoyang Li}, {and} \bibinfo{person}{Qingsong Wen}.} \bibinfo{year}{2024}\natexlab{}.
\newblock \showarticletitle{Foundation Models for Education: Promises and Prospects}.
\newblock \bibinfo{journal}{\emph{arXiv preprint arXiv:2405.10959}} (\bibinfo{year}{2024}).
\newblock


\bibitem[Yeti{\c{s}}tiren et~al\mbox{.}(2023)]%
        {yeticstiren2023evaluating}
\bibfield{author}{\bibinfo{person}{Burak Yeti{\c{s}}tiren}, \bibinfo{person}{I{\c{s}}{\i}k {\"O}zsoy}, \bibinfo{person}{Miray Ayerdem}, {and} \bibinfo{person}{Eray T{\"u}z{\"u}n}.} \bibinfo{year}{2023}\natexlab{}.
\newblock \showarticletitle{Evaluating the code quality of ai-assisted code generation tools: An empirical study on github copilot, amazon codewhisperer, and chatgpt}.
\newblock \bibinfo{journal}{\emph{arXiv preprint arXiv:2304.10778}} (\bibinfo{year}{2023}).
\newblock


\bibitem[Zhong et~al\mbox{.}(2024)]%
        {zhong2024influences}
\bibfield{author}{\bibinfo{person}{Tianlong Zhong}, \bibinfo{person}{Gaoxia Zhu}, \bibinfo{person}{Chenyu Hou}, \bibinfo{person}{Yuhan Wang}, {and} \bibinfo{person}{Xiuyi Fan}.} \bibinfo{year}{2024}\natexlab{}.
\newblock \showarticletitle{The influences of ChatGPT on undergraduate students’ demonstrated and perceived interdisciplinary learning}.
\newblock \bibinfo{journal}{\emph{Education and Information Technologies}} (\bibinfo{year}{2024}), \bibinfo{pages}{1--27}.
\newblock


\end{thebibliography}

\end{document}